\documentclass[12pt]{iopart}
\usepackage{iopams}
\begin{document}

\def\be{\begin{equation}}
\def\ee{\end{equation}}
\def\bey{\begin{eqnarray}}
\def\eey{\end{eqnarray}}
\def\hl{h(l)}
\def\phizero{\tau(0)}
\def\varphizero{\varphi(0)}
\def\psizero{\psi(0)}
\def\hzero{h(0)}
\def\phil{\tau(l)}
\def\varphil{\varphi(l)}
\def\psil{\psi(l)}
\def\yT{{y_\tau}}
\def\yvarphi{{y_\varphi}}
\def\yH{{y_h}}
\def\ypsideux{y_{\psi^2}}
\def\ypsitrois{y_{\psi^3}}
\def\yTpsi{y_{\tau\psi}}
\def\yvarphipsi{y_{\varphi\psi}}
\def\yHpsi{y_{h \psi}}
\def\Eq#1{Eq.~(\ref{#1})}
\def\Eqs#1#2{Eqs.~(\ref{#1}-\ref{#2})}
\def\EqsAnd#1#2{Eqs.~(\ref{#1}) and (\ref{#2})}
\def\tmod{|\tau|}
\def\gzero{g_0}
\def\gl{g}
\def\yg{y_g}
\def\yTg{y_{\phi g}}
\def\nuH{\nu_h}
\def\muH{\mu_h}

\title[The two-dimensional 4-state Potts model in a magnetic field]{The two-dimensional 4-state Potts model in a magnetic field}

\author{B. Berche$^{1}$, P. Butera$^{2}$, L.N. Shchur$^{3}$}

\address{$^{1}$ Statistical Physics Group, 
	Institut Jean Lamour\footnote{Laboratoire associ\'e au CNRS UMR 7198},
 	CNRS -- Universit\'{e} de Lorraine - Campus de Nancy, B.P. 70239, 
	F - 54506 Vand{\oe}uvre l\`es Nancy Cedex, France, EU}

\address{$^{2}$ Istituto Nazionale di Fisica Nucleare,
    	Sezione di Milano-Bicocca,\\
    	Piazza delle Scienze 3, 20126, Milano, Italia, EU\\
	  and Physics Department of Milano-Bicocca University}

\address{$^{3}$ Landau Institute for Theoretical Physics,
  	Russian Academy of Sciences, \\
    	Chernogolovka 142432, Russia}

\begin{abstract}
We present a solution of the non-linear renormalization group 
equations leading to the dominant and subdominant singular behaviours of 
physical quantities (free energy density, correlation length, 
internal energy, 
specific heat, magnetization, susceptibility and magnetocaloric coefficient) at the critical temperature
in a non-vanishing magnetic field. 
The solutions i) lead to exact cancellation of logarithmic corrections in universal amplitude ratios and ii) prove 
recently proposed relations among logarithmic exponents.
\end{abstract}



\eads{\mailto{berche@ijl.nancy-universite.fr}}

\maketitle

\section{Introduction, notations and definitions}

In a recent paper on the amplitude combinations in the 4-state Potts
model~\cite{ShchurBercheButera09}, we have provided in an appendix a
detailed analysis of the non-linear Salas and Sokal Renormalization
Group (RG) equations~\cite{SalasSokal97}, leading in particular to the
critical behavior of densities of the free energy, the internal
energy, the specific heat, the magnetization and the susceptibility in
zero external magnetic field.

A similar analysis, that was not written in
Ref.~\cite{ShchurBercheButera09}, would allow the calculation of the
same quantities (including also the correlation length and magnetocaloric coefficient) {\em at the
  critical temperature in a non-zero magnetic field}.
Recently this problem was considered in Ref.~\cite{co-workers} and
this motivated us to revisit our approach and observe that it can i)
complete the RG determination of all critical exponents, ii)
demonstrate the exact cancellation of logarithmic corrections in
universal amplitude ratios to all orders and iii) automatically lead
to the scaling laws introduced by Kenna, Johnston and Janke among the 
logarithmic correction
exponents~\cite{KennaJohnstonJanke1,KennaJohnstonJanke2,Kenna12}.

Let us remind first the standard way of deriving universal combinations of
critical amplitudes.  For this purpose, we will illustrate the case of
the non-trivial ratios $R_C^\pm=A_\pm\Gamma_\pm/B_-^2$ and
$R_\chi^\pm=\Gamma_\pm D_cB_-^{\delta-1}$ which connect the amplitudes in
the high and low temperature phases or at the critical temperature in presence of a
magnetic field. We will see that the second ratio,
connecting simultaneously temperature scaling and magnetic field scaling will be source of difficulties
when logarithmic corrections will be taken into account.

 A basic hypothesis in the theory of critical phenomena which relies on
the RG analysis is the homogeneity assumption for the singular part of
the free energy density~\cite{PatashinskiPokrovski,Kadanoff} \be
f_s(\tau,h)=b^{-D}F_\pm(\kappa_\tau b^\yT|\tau|,\kappa_h b^\yH |h|)
\ee where $D$ is the space dimension, $b$ is the rescaling factor,
$\tau$ and $h$ are the relevant thermal and magnetic fields with the
corresponding RG eigenvalues $\yT $ and $\yH $, $F_\pm(x,y)$ is a
universal function of its arguments $x$ and $y$, $\pm$ stands for
$T>T_c$ and $T<T_c$, and $\kappa_\tau$ and $\kappa_h$ are non-universal 
metric factors (which depend e.g. on the lattice symmetry
at a given space dimension).  The critical behaviors of the
magnetization, the susceptibility and the specific heat then follow by
taking derivatives with respect to $\tau$ or $h$,
\begin{eqnarray}
M_c(0,h)&=\kappa_hb^{-D+\yH}\partial_hF_c(x,y)|_{x=0},&\tau=0,\ h \to 0\label{eq-1a}\\
M_-(\tau,0)&=\kappa_h b^{-D+\yH}\partial_hF_-(x,y)|_{y=0},&\tau\to 0^-,\ h=0\label{eq-1}\\
\chi_\pm(\tau,h)&=\kappa_h^2 b^{-D+2\yH}\partial_h^2F_\pm(x,y),&\tau\to 0,\ h\to 0\label{eq-1c}\\
C_\pm(\tau,h)&=\kappa_\tau^2 b^{-D+2\yT}\partial_\tau^2F_\pm(x,y),&\tau\to 0,\ h\to 0\label{eq-1e}.
\end{eqnarray}
The definition of the critical exponents according to the standard terminology
 follows from the elimination of 
$x$ and $y$ dependence {\it at the critical temperature, $b=(\kappa_h|h|)^{-1/\yH},$ $\tau=0$, $h\to 0$}:
\begin{eqnarray}
&M_c(h)=D_c^{-1/\delta}|h|^{1/\delta}, 
    &\delta=\frac{\yH}{D-\yH},
    \quad D_c^{-1/\delta}=\kappa_h^{1+1/\delta}\partial_hF_c(0,1),\label{defBc}
\end{eqnarray}
or {\it in zero magnetic field, $b=(\kappa_\tau|\tau|)^{-1/\yT}$, $\tau\to 0$, $h=0$}:
\begin{eqnarray}
&M_-(\tau)=B_-(-\tau)^\beta, 
    &\beta=\frac{D-\yH}{\yT}, 
    \quad B_-=\kappa_h\kappa_\tau^\beta\partial_hF_-(1,0), 
    \  \tau\to 0^-\label{defB}\\
&\chi_\pm(\tau)=\Gamma_\pm|\tau|^{-\gamma}, 
    &\gamma=\frac{2\yH-D}{\yT},
    \quad\Gamma_\pm =\kappa_h^2\kappa_\tau^{-\gamma} \partial_h^2F_\pm(1,0),\label{defGamma}\\
&C_\pm(\tau)=\frac{A_\pm}\alpha|\tau|^{-\alpha}, 
    &\alpha=\frac{2\yT-D}{\yT}, 
    \quad\frac{A_\pm}{\alpha}=\kappa_\tau^{2-\alpha} \partial_\tau^2F_\pm(1,0)\label{defA}.
\end{eqnarray}
The subscript $c$, e.g. in $F_c(0,y)$, specifies that the function is evaluated at the critical temperature $\tau = 0$.

The amplitudes are clearly non-universal quantities, but elimination of all non-universal metric factors
is possible by forming
convenient combinations of these amplitudes which are universal. Notice that $\kappa_h$ disappears from the ratio $\chi_\pm(\tau)/M_-^2(\tau)$,
then, exploiting the Rushbrooke scaling law $\alpha+2\beta+\gamma=2$, we multiply this quantity by $C_\pm(\tau)$
to  eliminate also $\kappa_\tau$ and finally we obtain the function 
\begin{equation}
R_C^\pm(\tau)=|\tau|^2 C_\pm(\tau)\chi_\pm(\tau)/M_-^2(\tau)\label{eq-RC}\end{equation} 
tending to the universal quantity $\partial_\tau^2F_\pm(1,0)\partial_h^2F_\pm(1,0)/(\partial_hF_\pm(1,0))^2$=
$A_\pm\Gamma_\pm/\alpha B_-^2$ which establishes the universality of this 
 combination of critical amplitudes. The last equality above follows from the definitions of amplitudes in equations~(\ref{defB}), 
(\ref{defGamma}) and (\ref{defA}).
Clearly, a universal combination is associated to a scaling law, Rushbrooke scaling law in the present case.

But this is not the whole story, since one knows that in some cases 
 (as for
 the 4$-$state Potts model in two dimensions) logarithmic corrections 
occur which involve ``hat exponents''~\cite{CardyNauenbergScalapino80,Kenna12}, 
e.g. $M_-(\tau)=B_-|\tau|^\beta(-\ln|\tau|)^{\hat\beta}$, $\chi_\pm(\tau)=\Gamma_\pm|\tau|^{-\gamma}(-\ln|\tau|)^{\hat\gamma}$ and
$C_\pm(\tau)=\frac{A_\pm}{\alpha}|\tau|^{-\alpha}(-\ln|\tau|)^{\hat\alpha}$. The combinations $R_C^\pm(\tau)$ in equation~(\ref{eq-RC}) now
tends towards
\be R_C^\pm(\tau)\to\frac{A_\pm\Gamma_\pm}{\alpha B_-^2}(-\ln|\tau|)^{\hat\alpha-2\hat\beta+\hat\gamma},
\ee
and, provided that no other log-term appears, we have to impose the scaling relation \be\hat\alpha-2\hat\beta+\hat\gamma=0\label{eqhatscalingrushbrooke}\ee   
among the exponents describing
the logarithmic corrections
  in order to still guarantee the universality of the combinations $A_\pm\Gamma_\pm/{B_-^2}$.

The same line of reasoning for the other combinations considered, $R_\chi^\pm$, is less obvious (and this is the reason why we have
chosen this ratio to illustrate our purpose). It is easy to show that in the absence
of log-corrections, thanks to the Widom
scaling relation
$\gamma=\beta(\delta-1)$,
the function 
\begin{equation}R^\pm_\chi(\tau,h)=|h|\chi_\pm(\tau)M_-^{\delta-1}(\tau)M_c^{-\delta}(h)\label{eq-RChi}\end{equation}  tends to the universal quantity $\partial_h^2F_\pm(1,0)(\partial_hF_-(1,0))^{\delta-1}(\partial_hF_c(0,1))^{-\delta}$. It follows, using equations~(\ref{defBc}), 
(\ref{defB}) and (\ref{defGamma}), that the  
combinations $\Gamma_\pm D_cB_-^{\delta-1}$ are universal. On the other hand, when logarithmic
corrections occur,  one obtains the limiting behavior 
\be R_\chi(\tau,h)\to \Gamma_\pm D_c B_-^{\delta-1}(-\ln|\tau|)^{\hat\gamma+\hat\beta(\delta-1)}
(-\ln|h|)^{-\delta\hat\delta}\label{eqFive}
\ee
from which one would be tempted to conclude {\em erroneously} that $\hat\gamma+\hat\beta(\delta-1)=0$ and $\delta\hat\delta=0$.
This is not correct, as we will see later, because of an interplay between the two types of logarithms in $|\tau|$ and in $|h|$
expressed by equation~(\ref{eqtauh}) below.
We thus have to improve the analysis presented in this introductory section.

\section{Renormalization Group analysis}

In order to solve the problem, we provide below a re-examination of
the RG derivation of all scaling quantities, including now the
dependence on an external magnetic field along the critical isotherm.
Let us first remind the standard definitions of some exponent
combinations which will occur below: $\alpha_c=\alpha/\beta\delta,
\beta_c=\beta/\beta\delta$, $\gamma_c=\gamma/\beta\delta$
$\nu_c=\nu/\beta\delta$, $\epsilon_c=1-\alpha_c$.  For the critical
amplitudes we use the notations of
Refs.~\cite{PrivmanHohenbergAharony91,Kenna12}.
\begin{eqnarray}
 h=0,\ \tau\to 0^\pm, & \tau=0,\ h\to 0^\pm, \\
 f_s(\tau,\psi)=F_\pm|\tau|^{2-\alpha}(-\ln|\tau|)^{\hat\alpha}, & f_s(h,\psi)=F_c|h|^{1+1/\delta}(-\ln|h|)^{\hat\delta}, \\
 M(\tau,\psi)=B_-|\tau|^{\beta}(-\ln|\tau|)^{\hat\beta},
 & M(h,\psi)=D_c^{-1/\delta}|h|^{1/\delta}(-\ln|h|)^{\hat\delta}, \\
 E(\tau,\psi)=\frac{A_\pm}{\alpha(1-\alpha)}|\tau|^{1-\alpha}(-\ln|\tau|)^{\hat\alpha}, & E(h,\psi)=E_c|h|^{\epsilon_c}(-\ln|h|)^{\hat\epsilon_c}, \\
 \chi(\tau,\psi)=\Gamma_\pm|\tau|^{-\gamma}(-\ln|\tau|)^{\hat\gamma}, & \chi(h,\psi)=\Gamma_c|h|^{1/\delta-1}(-\ln|h|)^{\hat\delta}, \\
 C(\tau,\psi)=\frac{A_\pm}{\alpha}|\tau|^{-\alpha}(-\ln|\tau|)^{\hat\alpha}, & C(h,\psi)=\frac{A_c}{\alpha_c}|h|^{-\alpha_c}(-\ln|h|)^{\hat\alpha_c}, \\
 m_T(\tau,\psi)=m_\pm|\tau|^{\beta-1}(-\ln|\tau|)^{\hat\beta},  & m_T(h,\psi)=m_c|h|^{\epsilon_c-1}(-\ln|h|)^{\hat\epsilon_c},\\ 
 \xi(\tau,\psi)=\xi_\pm|\tau|^{-\nu}(-\ln|\tau|)^{\hat\nu}, & \xi(h,\psi)=\xi_c|h|^{-\nu_c}(-\ln|h|)^{\hat\nu_c}. 
\end{eqnarray}
In these expressions, $\psi$ denotes an irrelevant field, the role of which is discussed below.
The magnetocaloric coefficient $m_T$,  is often measured in experiments - since it is more singular that the magnetization - but 
 is not an independent quantity.

The non-linear Salas and Sokal RG equations~\cite{SalasSokal97,CardyNauenbergScalapino80,NauenbergScalapino80} for the 
relevant thermal and magnetic
fields $\tau$ and $h$ and
the marginal dilution field $\psi$, are given by
\begin{eqnarray}
    &&\frac{d\tau }{d l}=(\yT +\yTpsi \psi )\tau
        ,\label{ap-eq2}\\
    &&\frac{dh }{d l}=(\yH +\yHpsi \psi)h,
        \label{ap-eq3}\\
    &&\frac{d\psi  }{d l}=g(\psi).
        \label{ap-eq1}
\end{eqnarray}
with 
$l=\ln b$.
The fixed point is at $\tau=h=0$. Starting from initial conditions
$\phizero$, $\hzero$, the relevant fields $\tau$ and $h$ grow exponentially 
with $l$, 
and their behaviours follow from the renormalization 
flow from $\phizero\sim\tau$, $\hzero\sim h$ in the vicinity of the critical point up to some $\phil=O(1)$,
$\hl=O(1)$ outside the
critical region. 
The function $g(\psi)$ is smooth and  may be expanded in powers of $\psi$,
$g(\psi)=\ypsideux \psi^2+\ypsitrois\psi^3+\dots$.
The dilution field $\psi$ being marginal for $q=4$ in two dimensions, 
there is no linear term in $g(\psi)$ and along the RG flow $\psil$
remains of order $O(\psizero)$ and $\psizero$ is negative,
$|\psizero|=O(1)$. 
The first term in the function $g(\psi)$ was first considered by Nauenberg and
Scalapino~\cite{NauenbergScalapino80}, and later by
Cardy, Nauenberg and Scalapino~\cite{CardyNauenbergScalapino80}, and 
the second term was introduced
by Salas and Sokal~\cite{SalasSokal97}.
The parameters are known and
take the values  $\yTpsi =3/(4\pi)$,
$\yHpsi =1/(16\pi)$, $\ypsideux =1/\pi$ and
$\ypsitrois =-1/(2\pi^2)$~\cite{CardyNauenbergScalapino80,SalasSokal97}, while
the relevant scaling dimensions are
$\yT =3/2$ and $\yH =15/8$.

In zero magnetic field, under a change of the length
scale, the singular part of the free energy density and the correlation
length transform
according to 
\bey
    f_s(\phizero,0,\psizero)&=&e^{-Dl}f_s(\phil,0,\psil),\label{ap-eq4}\\
    \xi(\phizero,0,\psizero)&=&e^{l}\xi(\phil,0,\psil),\label{ap-eqxi}
\eey 
where $D=2$ is the space dimension. 
Similarly, at $\tau=0$, the
dependence on the magnetic field obeys
\bey
    f_s(0,\hzero,\psizero)&=&e^{-Dl}f_s(0,\hl,\psil),\label{ap-eq4h}\\
    \xi(0,\hzero,\psizero)&=&e^{l}\xi(0,\hl,\psil).\label{ap-eqxih}
\eey

The thermal behaviour in zero magnetic field is obtained by solving 
\EqsAnd{ap-eq2}{ap-eq1}, while the dependence on the magnetic field 
 along the critical isotherm follows from \EqsAnd{ap-eq3}{ap-eq1}. 
The two sets of equations have exactly the same structure. We will therefore
use a common notation $\varphi$ for the relevant scaling field ($\tau$ or $h$).
Starting in the vicinity of the critical point at $\varphi(0)=\varphi$, the field grows under renormalization as $\varphi'=\varphi(l)=\varphi e^{\yvarphi l}+\ \!$corrections$\ \!\sim b^\yvarphi\varphi\ \!+\ \!$corrections.
This provides the leading singularities in Eqs~(\ref{ap-eq4}) to (\ref{ap-eqxih}), and the homogeneity assumption
approximately takes the usual form
$f_s(\varphi,\psizero)=b^{-D}f_s(b^\yvarphi\varphi,\psil)$ and $\xi(\varphi,\psizero)=b\xi(b^\yvarphi\varphi,\psil)$.
The correction terms
will be responsible for the appearance of logarithms in all physical quantities and the purpose of the present paper
is essentially to 
analyse in detail the role of the corrections. 
We shall discuss in particular the relations among the ``{hat}-exponents'' 
introduced by Kenna, Johnston and Janke which are still known only through
the scaling laws derived by these authors~\cite{Kenna12}.

\Eq{ap-eq2} (or \Eq{ap-eq3}) leads to
\be\int_0^l\frac{d\varphi}\varphi
=\ln\frac{\varphil}{\varphizero}={\rm const}+\ln\frac 1{|\varphi|}
=\yvarphi l+\yvarphipsi \int_0^l\psi dl
        ,\label{ap-eqaa}\ee 
where the last integral is
obtained from \Eq{ap-eq1} rewritten as
\be\psi dl=\frac 1{\ypsideux}\left(\frac 1{\psi}-\frac{\ypsitrois}
{\ypsideux+\ypsitrois\psi}\right)d\psi,\ee 
thus
\be\int_0^l\psi dl=\frac 1{\ypsideux }\ln \left(\frac{\psil}{\psizero}
\frac{\ypsideux+\ypsitrois\psizero}{\ypsideux+\ypsitrois\psil}\right).
 \label{ap-eqbb}\ee
Combining \Eq{ap-eqaa} and \Eq{ap-eqbb} 
we get 
\be
    l={\rm const}-\frac 1{\yvarphi }\ln |\varphi|+\frac{\yvarphipsi }{\yvarphi \ypsideux }
    \ln \frac{\psizero}{\psil}
\frac{\ypsideux+\ypsitrois\psil}{\ypsideux+\ypsitrois\psizero}.
    \label{ap-eq5}
\ee
Apart from the leading $\varphi-$dependence, all logarithmic corrections which
occur in the $4-$state Potts model are encoded in
the dilution field dependence. The ubiquitous combination where it appears
 is conveniently denoted as
\be\zeta=\frac{\psil}{\psizero}
\frac{\ypsideux+\ypsitrois\psizero}{\ypsideux+\ypsitrois\psil}\label{eq-zeta}
\ee
and 
 remembering that $\varphi$ is either the reduced
temperature $\tau$, or the external magnetic field $h$, 
\bey
    f_s(\varphi,\psizero)&=&{\rm const}\times|\varphi|^{D/\yvarphi}
	\zeta^{{D\yvarphipsi }/{\yvarphi \ypsideux }}
	\\
    \xi(\varphi,\psizero)&=&{\rm const}\times|\varphi|^{-1/\yvarphi}
	\zeta^{{-\yvarphipsi }/{\yvarphi \ypsideux }}
\eey

When we specify  $\varphi$, we obtain from Eq.~(\ref{ap-eq5}) the functional similarity
\begin{equation}|\tau|^\nu\zeta^\mu\propto|h|^{\nu_c}\zeta^{\mu_c}\label{eqtauh}\end{equation}
where it is convenient to denote
$\nu={1}/{\yT}={2}/{3},\ \nu_c={1}/{\yH}={8}/{15},\ \mu={\yTpsi}/{\yT\ypsideux}={1}/{2},
\ \mu_c={\yHpsi}/{\yH\ypsideux}={1}/{30}$. 
The free energy density is then written as $f_s(\tau,\psi)\sim|\tau|^{D\nu}\zeta^{D\mu}$ in zero magnetic field while
the field dependence along the critical isotherm 
is given by $f_s(h,\psi)\sim|h|^{D\nu_c}\zeta^{D\mu_c}$.
The other thermodynamic properties follow by derivation with respect to
the scaling fields, e.g.
$E(\tau,\psi)=\frac \partial{\partial \tau}f_s(\tau,\psi)$ which leads to either 
$E(\tau,\psi)\sim|\tau|^{D\nu-1}\zeta^{D\mu}$ when the magnetic field tends to zero,
or to $E(h,\psi)\sim|h|^{D\nu_c}\zeta^{D\mu_c}|\tau|^{-1}$ 
in the vicinity of the the critical isotherm. 
Using Eq.~(\ref{eqtauh}) when needed, and specifying either $h=0$ or $\tau=0$, 
we may now collect the following expressions,
\begin{eqnarray}
& h=0,\ \tau\to 0^\pm, & \tau=0,\ h\to 0^\pm, \\
\noalign{\vskip1mm}
& f_s(\tau,\psi)\sim |\tau|^{D\nu}\zeta^{D\mu}, & f_s(h,\psi)\sim|h|^{D\nu_c}\zeta^{D\mu_c}, \label{eq-tauhzetaf}\\
& M(\tau,\psi) \sim |\tau|^{D\nu-\frac{\nu}{\nu_c}}\zeta^{D\mu-\frac{\mu-\mu_c}{\nu_c}}, & M(h,\psi)\sim |h|^{D\nu_c-1}\zeta^{D\mu_c}, \\
& E(\tau,\psi) \sim |\tau|^{D\nu-1}\zeta^{D\mu}, & E(h,\psi)\sim |h|^{D\nu_c-\frac{\nu_c}{\nu}}\zeta^{D\mu_c-\frac{\mu_c-\mu}{\nu}}, \\
& \chi(\tau,\psi) \sim  |\tau|^{D\nu-2\frac{\nu}{\nu_c}}\zeta^{D\mu-2\frac{\mu-\mu_c}{\nu_c}}, & \chi(h,\psi)\sim |h|^{D\nu_c-2}\zeta^{D\mu_c},\\
& C(\tau,\psi) \sim  |\tau|^{D\nu-2}\zeta^{D\mu}, & C(h,\psi)\sim |h|^{D\nu_c-2\frac{\nu_c}{\nu}}\zeta^{D\mu_c-2\frac{\mu_c-\mu}{\nu}},\\
& m_T(\tau,\psi) \sim  |\tau|^{D\nu-1-\frac{\nu}{\nu_c}}\zeta^{D\mu-\frac{\mu-\mu_c}{\nu_c}}, & m_T(h,\psi) \sim |h|^{D\nu_c-1-\frac{\nu_c}{\nu}} \zeta^{D\mu_c-\frac{\mu_c-\mu}{\nu}},\\
& \xi(\tau,\psi) \sim |\tau|^{-\nu}\zeta^{-\mu}, & \xi(h,\psi)\sim |h|^{-\nu_c}\zeta^{-\mu_c} .
\label{eq-tauhzeta}
\end{eqnarray}
The values of the leading exponents follow directly, 
\be\begin{array}{lll}
\alpha=2-D\nu=\frac 23, && \alpha_c=2\frac{\nu_c}{\nu}-\frac D{\nu_c}=\frac 8{15},\\
\beta=D\nu-\frac{\nu}{\nu_c}=\frac 1{12}, && \delta = \frac 1{D\nu_c-1}=15, \\
\gamma = 2\frac\nu{\nu_c}-D\nu=\frac 76, && \epsilon_c = D\nu_c-\frac{\nu_c}\nu=\frac4{15},\\
\nu=\frac 23, && \nu_c = \frac 8{15}.\\
\end{array}\label{eq-exponents}
\ee

\section{Exponents of logarithmic corrections and scaling relations among them}

We now want to explore the values of the ``{hat exponents}'' and the link to universal combinations of critical amplitudes. 
The particular form taken by the function $\zeta$ follows from
the solution of \Eq{ap-eq1}, combined to \Eq{ap-eq5} iterated at the 
convenient level of approximation   
(see Appendix of Ref.~\cite{ShchurBercheButera09} for details and 
Refs.~\cite{DelfinoCardy98,DelfinoBarkemaCardy00,CaselleTateoVinci99,EntingGuttmann03,ShchurBercheButera08,BercheButeraShchur10,BBJS09} for different levels of approximation).
Keeping only the leading logarithmic behavior for 
the present context,  expression (\ref{eq-zeta}) 
simply yields 
\be \zeta\sim (-\ln|\varphi|)^{-1}(1+{\rm corrections})
\ee
and the exponents of all logarithmic corrections are directly read in Eqs.~(\ref{eq-tauhzetaf}-\ref{eq-tauhzeta}):
\be\begin{array}{lll}
\hat\alpha=-D\mu=-1, && \hat\alpha_c=2\frac{\mu_c-\mu}\nu-D\nu_c=-\frac{22}{15},\\
\hat\beta=\frac{\mu-\mu_c}{\nu_c}-D\mu=-\frac 18, &&\hat\delta=-D\mu_c=-\frac 1{15},\\
\hat\gamma=2\frac{\mu-\mu_c}{\nu_c}-D\mu=\frac 34, && \hat\epsilon_c=\frac{\mu_c-\mu}\nu-D\mu_c=-\frac {23}{30},\\
\hat\nu=\mu=\frac 12, && \hat\nu_c=\mu_c=\frac 1{30}.\\
\end{array}\label{eq-exponents}
\ee

What appears extremely useful in these expressions is that when defining
appropriate effective ratios, the dependence on the quantity
$\zeta$ cancels, due to the scaling
relations among the critical exponents. This quantity $\zeta$
is precisely the only
one where the log terms are hidden, and thus
we may infer that not only the leading log terms, but all the log terms hidden
in the dependence on the marginal dilution field disappear in the conveniently
defined effective ratios.
For example in effective ratios like those considered in the introduction in equations~(\ref{eq-RC}) and (\ref{eq-RChi}),
\be R_C^\pm(\tau)=\tau^{2}\frac{C_\pm(\tau,\zeta)\chi_\pm(\tau,\zeta)}{M_-^2(\tau,\zeta)},\ee
\be R_\chi^\pm(\tau,h)=|h|\chi_\pm(\tau,\zeta)M_-^{\delta -1}(\tau,\zeta)M_c^{-\delta}(h,\zeta)
\label{rat-app},\ee 
all corrections to scaling coming from the
variable $\zeta$ disappear, provided that in addition to the scaling
relation~(\ref{eqhatscalingrushbrooke}), 
one more scaling
relation is also satisfied \be
\hat\gamma+(\delta-1)\hat\beta-\delta\hat\delta=0.\label{eqhatscalingbis}\ee
The two scaling laws~(\ref{eqhatscalingrushbrooke}) and  ~(\ref{eqhatscalingbis}) are verified by the values of the  ``hat exponents'' of the 4-state Potts
model given in equations~(\ref{eq-exponents}).

This solves the problem of the cancellation of the
logarithmic corrections identified in Eq.~(\ref{eqFive}) and the
approach is easily extended to the other universal combinations of
critical amplitudes. In conclusion this approach
 provides both the scaling relations among the leading exponents and those among the exponents of the logarithmic corrections.


\section*{Acknowledgements}
We thank Ralph Kenna for stimulating discussions on scaling relations and related phenomena.

\vskip1cm

\def\paper#1#2#3#4#5{
	#1, {\it #2}\ {\bf #3}, #4 (#5).}

\end{document}